\documentclass[traditabstract]{aa}  

\usepackage{amsmath}
\usepackage{amssymb}
\usepackage{natbib}

\usepackage{graphicx}
\usepackage{comment}
\usepackage{txfonts}
\usepackage{ctable}
\usepackage{hyperref}
\usepackage{xspace}
\usepackage{epstopdf, epsfig}
\usepackage{tabularx}
\usepackage{threeparttable}
\usepackage{caption}
\usepackage{subcaption}
\usepackage{hyperref}

\def\hess{H.E.S.S.}
\def\chandra{\textit{Chandra}}
\def\xmm{\textit{XMM-Newton}}
\def\suzaku{\textit{Suzaku}}

\def\rosat{\textit{ROSAT}}

\def\fermi{\textit{Fermi}-LAT}

\def\velajr{Vela Jr.}

\newcommand{\mpo}[1]{#1}
\newcommand{\rob}[1]{#1}

\newcommand{\is}[1]{#1}
\newcommand{\isr}[1]{#1}
\newcommand{\isrr}[1]{#1}

\begin{document}




\title{Modeling of the spatially resolved non-thermal emission from the \velajr\ supernova remnant}


\author{Iurii Sushch \inst{1,2}
\and Robert Brose \inst{1,3}
\and Martin Pohl \inst{1,3}
}
\institute{DESY, D-15738 Zeuthen, Germany  
  \and
  Astronomical Observatory of Ivan Franko National University of L'viv, vul. Kyryla i Methodia, 8, L'viv 79005, Ukraine
  \and 
  Institute of Physics and Astronomy, University of Potsdam, 14476, Potsdam, Germany
}
\date{Received 22 February, 2018; accepted 26 July, 2018}

\abstract{Vela Jr. (RX J0852.0$-$4622) is one of just a few known supernova remnants
(SNRs) with \isr{a} resolved shell across the whole electromagnetic spectrum
from radio to very-high-energy ($>100$ GeV; VHE) gamma-rays. Its proximity
and large size allow for detailed spatially resolved observations of the
source making Vela Jr. one of the primary sources used for the study of
particle acceleration and emission mechanisms in SNRs. High-resolution
X-ray observations reveal \rob{a} steepening of the spectrum toward the interior
of the remnant. In this study we aim for a self-consistent radiation model
of Vela Jr. which at the same time would explain the broadband emission from
the source and its intensity distribution. We solve the full particle
transport equation combined with the high-resolution 1D hydrodynamic
simulations (using Pluto code) and subsequently calculate the radiation from
the remnant. \is{\rob{The e}quations are solved in the test particle regime.} We test two models for the magnetic field profile downstream of the shock: damped magnetic field which accounts for the damping of strong magnetic
turbulence downstream, and transported magnetic field. Neither of these scenarios can fully explain the observed radial dependence of the X-ray spectrum under spherical symmetry. We show, however, that the softening of the spectrum and the X-ray intensity profile can be explained under the assumption that the emission is enhanced within a cone.
}

\keywords{radiation mechanisms: non-thermal -- acceleration of particles -- cosmic rays -- shock waves -- ISM: supernova remnants -- X-rays: individuals: Vela Jr. (RX J$0852.0-4622$)}







\authorrunning{Sushch et al.}
\titlerunning{Modeling of the \velajr\ SNR}
\maketitle


\section{Introduction}

Supernova remnants (SNRs) are widely considered to be the main candidates for the acceleration sites of Galactic cosmic rays (CRs)\isr{,} which places them among the most interesting and most studied astrophysical objects. \velajr\ is one of just a few SNRs with detected non-thermal emission and \rob{a} well resolved shell-like morphology across the whole electromagnetic spectrum from radio to very-high-energy ($\gtrsim100$ GeV; VHE) gamma-rays. Its proximity, large size and strong non-thermal emission (specifically from the north-western rim of the remnant) allows for detailed spatially resolved observations of the source both in X-rays and at TeV energies making it one of the prime sources for studies of particle acceleration and radiation mechanisms in SNRs. Despite a rather late discovery of the source by the \rosat\ satellite at X-ray energies above $\sim1.3\,$keV \citep{1998Natur.396..141A} owing to its location in a very crowded region with the bright and large Vela\,SNR in the foreground, it became one of the most observed remnants across the electromagnetic spectrum. 

\velajr\ was one of the first sources detected in the VHE range \citep{paper:vjr_hess_paper1} and benefits from a large amount of data collected with the \hess\ telescopes since 2004. It was also detected by \fermi\ in the GeV energy range \citep{paper:vjr_fermi}. Both \fermi\ and \hess\ maps resolve a clear shell-like morphology of the remnant. The spectrum in the GeV range obtained by \fermi\ and the spectrum in the TeV range obtained by \hess\ connect remarkably well and can be nicely described by a power law with an exponential cutoff with the cut-off energy of $(6.7\pm1.2_{\rm stat}\pm1.2_{\rm syst})$\,TeV \citep{2016arXiv161101863H}.

It is still not clear which process is responsible for the gamma-ray emission from the remnant. Both hadronic (proton interactions with a subsequent $\pi^0$-decay) and leptonic (inverse Compton (IC) scattering of relativistic electrons on ambient radiation fields) scenarios can adequately reproduce the shape
of the gamma-ray spectrum \citep{2016arXiv161101863H}. 
A fit of the combined GeV-TeV spectrum with the IC emission from a parent electron population of the form of a power law with exponential cutoff in a pure leptonic scenario yields the electron spectral index of $2.33\pm0.03_{\rm stat}\pm0.33_{\rm syst}$ and the electron cutoff energy of $(27 \pm 1_{\rm stat} \pm 12_{\rm syst})$\,TeV. In the hadronic scenario, the fit of the observed spectrum results in a parent proton spectrum with the spectral index of $1.83\pm0.02_{\rm stat}\pm0.11_{\rm syst}$ and \rob{a} cutoff energy of $(55 \pm 6_{\rm stat} \pm 13_{\rm syst})$\,TeV \citep{2016arXiv161101863H}. Both scenarios, however, face some difficulties. The hadronic scenario is supported by a good spatial correlation of the TeV emission with the distribution of the HI gas \citep{2017arXiv170807911F}, but the lack of thermal X-ray emission from the remnant places the upper limit on the ambient density at $\sim0.03$\,cm$^{-3}$ \citep{2001ApJ...548..814S}\isr{,} which makes the hadronic scenario rather implausible as it would require the total energy in protons to be comparable to the explosion energy. For RX\,J1713.7-3946, which shares a lot of observational properties with \velajr, it was suggested that the lack of thermal X-ray emission can still be explained within the hadronic scenario if the remnant is expanding in a medium with small dense clouds\isr{,} which survive the passage of the shock and play a role of target material for cosmic rays \citep[][]{2012ApJ...744...71I, 2014MNRAS.445L..70G}. However, this scenario also has some difficulties \citep{2015A&A...577A..12F}. A large mass in clouds required for a fit of the high energy spectrum in this scenario requires pre-existing \isr{clouds,} which have been unaffected by the stellar wind of the progenitor star, which is rather implausible as the stellar wind would form Kevin-Helmholtz and Rayleigh-Taylor instabilities destroying \rob{the} outer layers of \rob{the} clouds. This would be followed by \rob{the} formation of $\sim(1-10)\,$cm$^{-3}$ streams of gas\isr{,} which would strongly emit in X-rays when interacting with the forward shock \citep{2015A&A...577A..12F}. 

On the other hand high resolution X-ray observations reveal thin and bright filamentary structures in the north western (NW) part of the shell \citep{2005ApJ...632..294B}. These filaments are usually explained by fast synchrotron cooling in \rob{a} strongly amplified magnetic field \citep[$B\gtrsim100\,\mu$G; see e.g.][]{2009A&A...505..641B}. This strong magnetic field is in contradiction with the leptonic scenario of the gamma-ray emission as in this case a relatively weak field of $\sim10\,\mu$G is required to explain the overall X-ray emission from the remnant \citep{2016arXiv161101863H, 2013ApJ...767...20L}. However, it is also possible that X-ray filaments are limited by magnetic field damping downstream of the forward shock \citep{2005ApJ...626L.101P, 2012A&A...545A..47R}. In this scenario the magnetic field downstream of the shock can be weaker as the width of the filament is determined by the length scale at which strong magnetic turbulence formed upstream is damped downstream.

Recently a detailed analysis of the archival \xmm\ data revealed a gradual softening of the spectrum (from a photon index of 2.56 to a photon index of 2.96) in the north western rim towards the interior of the remnant \citep{2013A&A...551A.132K}. This softening was \isr{interpreted} as the \is{decrease of the maximum electron energy with time due to synchrotron cooling.} Using a simple spectral evolution model assuming that the downstream time evolution of the maximum electron energy is determined only by synchrotron losses, \citet{2013A&A...551A.132K} \isrr{found} that the softening of the spectrum can be explained by a rather low magnetic field of \isr{$\sim10$}\,$\mu$G. \isr{However, this requires an electron maximum energy of about $50\,$TeV which is higher than \mpo{that} suggested by \hess\ observations at TeV energies \citep{2016arXiv161101863H}}. More importantly, these calculations were done neglecting a projection effect which \isr{can potentially} strongly weaken the spatial variation of the spectrum \isr{depending on the 3D structure of the remnant}. \isr{It can also be shown that such a gradual softening of the spectrum can be similarly explained by magnetic field damping. Indeed, the cut-off \mpo{frequency} of the synchrotron spectrum decreases linearly with the decrease of the magnetic field strength hence the damping of the magnetic field downstream of the shock can potentially result in the similar spectral variation as for the synchrotron cooling. Therefore, in this work we test both scenarios properly accounting for the projection effect.}

Recent observations with \suzaku\ \citep{2016PASJ...68S..10T} also reveal a faint hard X-ray emission from the NW rim of the remnant. The spectrum in the energy range from 12 keV to 22 keV can be described with a power law with a spectral index of $3.15^{+1.18}_{-1.14}$. This result challenges the existence of a roll-off in the X-ray spectrum or at least suggests that the cut off of the electron spectrum is slower than exponential. However, the uncertainties on the spectral parameters are too large to draw strong conclusions.

It should also be mentioned that the southern part of the shell is coincident with a pulsar wind nebula (PWN) detected in X-rays with \xmm\ \citep{2013A&A...551A...7A}. \hess\ observations provide a hint that a fraction of up to 8\,\% of the total VHE emission from the \velajr\ region might be associated with this PWN \citep{2016arXiv161101863H}. If confirmed in future observation, e.g. with CTA, this should be taken into account for the broadband modeling of the emission.

In this paper we provide modeling of the non-thermal emission from the \velajr\ SNR focusing on explaining the morphology of the remnant and the constraints the observed morphology imposes. In this regard we consider two different profiles of the magnetic field downstream of the shock, damped magnetic field and transported magnetic field, to test which of \rob{the} two effects described above determines X-ray filaments and spectral softening observed in these filaments. For our simulation we use \rob{the} {\bf R}adiation {\bf A}cceleration {\bf T}ransport {\bf Pa}rallel {\bf C}ode (RATPaC) described in a number of previous papers \citep{2012APh....35..300T, 2013A&A...552A.102T, 2016A&A...593A..20B}. The code solves the time-dependent transport equation for cosmic rays in one dimension (1D) in the test particle regime and subsequently simulates the non-thermal radiation from accelerated particles. The shock evolution is simulated using 1D hydrodynamic simulations performed by the Pluto code \citep{2012ApJS..198....7M} which is the most recent addition to RATPaC. The calculation of the particle acceleration can also be coupled to the evolution of isotropic Alfv\'enic turbulence upstream of the shock \citep{2016A&A...593A..20B}.

\section{Modelling}

\subsection{\isr{Physical parameters of the SNR and its environment}}

We assume that \velajr\ was created in a core collapse Supernova (SN) explosion which is supported by the detection of the central compact object (CCO) AX\,J0851.9$-$4617.4\footnote{Also known as CXOU\,J085201.4$-$461753} close to the center of the remnant \citep{1998Natur.396..141A, 1999A&A...350..997A, 2001ApJ...548..814S}. However, it is not clear whether the association with the SNR is correct because no pulsations were detected and later it was also suggested that a potential CCO might be an unrelated planetary nebula \citep{2006A&A...449..243R}. Additionally, the absence of broad Ca\,II absorption lines in spectra of background stars together with the constraints set by the Ti$^{44}$ gamma-ray line also suggests a core collapse origin with SNe types of Ic or Ibc \citep{2010A&A...519A..86I}. In our hydrodynamic \isr{simulations (see below)} we assume \rob{an} explosion energy of $10^{51}$ erg and \rob{an} ejecta mass of $3\,M_\odot$ which roughly corresponds to the properties of \rob{a} Ibc type SN.  

The \isr{current} density of the ambient medium is constrained by the lack of the X-ray thermal emission which places the upper limit at $n_0<2.9\times10^{-2}\left(\frac{D}{1\,{\rm kpc}}\right)^{-1/2}f^{-1/2}\,{\rm cm^{-3}}$, where $D$ is the distance to the remnant and $f$ is the volume filling factor \isrr{of the X-ray emitting gas} \citep{2001ApJ...548..814S}. \isr{That limit applies to gas} with a temperature above 1\,keV, because at energies below 1\,keV the observed X-ray emission is dominated by the strong thermal emission from the Vela\,SNR.

The distance to the remnant and its age can be constrained by its angular size and its angular expansion rate measured by \citet{2008ApJ...678L..35K} and \citet{2015ApJ...798...82A} using \xmm\ and \chandra\ data respectively. \xmm\ data taken between 2001 and 2007 imply an expansion rate of $0.84^{\prime\prime} \pm 0.23^{\prime\prime}\,{\rm yr}^{-1}$ \citep{2008ApJ...678L..35K} while \chandra\ data from 2003 to 2008 suggests an expansion rate of $0.42^{\prime\prime} \pm 0.10^{\prime\prime}\,{\rm yr}^{-1}$ \citep{2015ApJ...798...82A}. The difference between the two measurements is somewhat puzzling because the data used to estimate the expansion rate were taken almost from the same region and almost over the same time period. In our simulations we use the measurement provided by \xmm\ to constrain the hydrodynamic model but also discuss how uncertainties on the expansion rate measurement can influence our results in Section\,\ref{anotherhydro}. The allowed range for the distance is $500-1000$\,pc. The lower limit is determined based on the requirement that the shock speed should be higher than $1000$\,km/s to effectively accelerate particles \citep{2015ApJ...798...82A}. The upper limit of $1$\,kpc is determined by the comparison of the detected $^{44}\mbox{Ti}$ line emission with evolution models for different types of supernova (SN) explosions \citep{2010A&A...519A..86I}. \isr{The comparison of the X-ray absorption column density to the distribution of $^{12}$CO and HI gas places \velajr\ in the foreground or at the interface of the Vela Molecular Ridge suggesting a similar upper limit of 900 pc \citep{2013A&A...551A...7A}.}

\subsection{Hydrodynamics}

\isr{The standard gasdynamical equations}
\begin{align}
\frac{\partial }{\partial t}\left( \begin{array}{c}
                                    \rho\\
				    \vec{m}\\
				    E
                                   \end{array}
 \right) + \nabla\left( \begin{array}{c}
                   \rho\vec{v}\\
		   \vec{mv} + P\vec{I}\\
		   (E+p)\vec{v} 
                  \end{array}
 \right)^T &= \left(\begin{array}{c}
                    0\\
		    0\\
		    0
                   \end{array}
 \right),\\
 \frac{\rho\vec{v}^2}{2}+\frac{P}{\gamma-1}  &= E 
\end{align}
are solved, where $\rho$ is the density of the thermal gas, $\vec{v}$ the plasma velocity, $\vec{m}=\vec{v}\rho$ the momentum density, $P$ the thermal pressure of the gas, $\vec{I}$ the unit tensor, and $E$ the total energy of the ideal gas with $\gamma=5/3$. We assume that the magnetic field is dynamically unimportant due to its low strength and the remnant not being in the radiative phase yet \citep{2016MNRAS.456.2343P}. The equations are solved in 1D for a spherical symmetry.

\isr{We initialize the ejecta profile by a plateau in density with the value $\rho_{\mathrm c}$ up to the radius $r_{\mathrm c}$ followed by a power-law distribution up to the ejecta-radius $R_{\mathrm{ej}}$:
\begin{align}
 \rho(r) &= \begin{cases}
             \rho_{\mathrm c}, & r<r_{\mathrm c},\\
             \rho_{\mathrm c}\left(\frac{r}{r_{\mathrm c}}\right)^{-n}, & r_{\mathrm c}\leq r \leq R_{\mathrm{ej}},\\
             \rob{\frac{\dot{M_\star}}{4\pi r^2 v_{wind}}}, & r>R_{\mathrm{ej}}.\\
            \end{cases}
	\label{gasdyn}
\end{align}
}
\isr{The exponent for the ejecta profile is \isrr{set to} $n=9$ for the core-collapse explosion. The velocity of the ejecta is defined as
\begin{equation}
v_{\mathrm{ej}}(r) = \frac{r}{T_{\mathrm{SN}}},
\end{equation}
where $T_{\mathrm{SN}}=1$\,yr is the initial time set for hydrodynamic simulations. Then for the assumed mass, $M_{\mathrm{ej}}$, and energy, $E_{\mathrm{ej}}$, of the ejecta, and defining the radius of the ejecta as multiple of $r_{\mathrm c}$, $R_{\mathrm{ej}} = xr_{\mathrm c}$, the initial conditions for simulations can be written as
\begin{align}
  r_c &=
  \left[\frac{10}{3}\frac{E_{ej}}{M_{ej}} \left(\frac{n-5}{n-3}\right) \left(\frac{1- \frac{3}{n} x^{3-n}}{1-\frac{5}{n} x^{5-n}} \right) \right]^{1/2} T_{SN}, \\
  \rho_c &= \frac{M_{ej}}{4\pi r_c^3}\frac{3(n-3)}{n} \isrr{\left(1- \frac{3}{n} x^{3-n}\right)^{-1}}, \\
  v_{\mathrm c} &= \frac{r_{\mathrm{c}}}{T_{\mathrm{SN}}}. 
\end{align}
}

\isr{The initial temperature is set to $10^4$~K everywhere and the initial pressure is calculated using the equation of state.} We used 200000 \isr{linearly distributed} grid-points and $14\,$pc grid-size for the hydro simulations. \isr{The heaviest ejecta bin contains about $0.1M_\odot$. For our simulations we set $x = \rob{3}$ resulting in $r_{\rm c} = 0.006$~pc and $R_{\rm ej} = 0.018$~pc.} 
To use the hydro-output for the particle acceleration, we had to resharpen the shock.    
We further assume \isrr{in Eq.~\ref{gasdyn}} that the SNR is expanding into a wind zone, i.e. the density radial distribution follows $n\propto r^{-2}$, and require the current upstream density to be lower than $0.03\,{\rm cm^{-3}}$. \isr{Simulations are stopped at the moment when the expansion rate of the remnant, i.e. \isrr{the ratio of the shock speed to the distance}, reaches the observed value (see the previous section).} 

\begin{figure}[ht!]
  \centering
  \resizebox{\hsize}{!}{\includegraphics{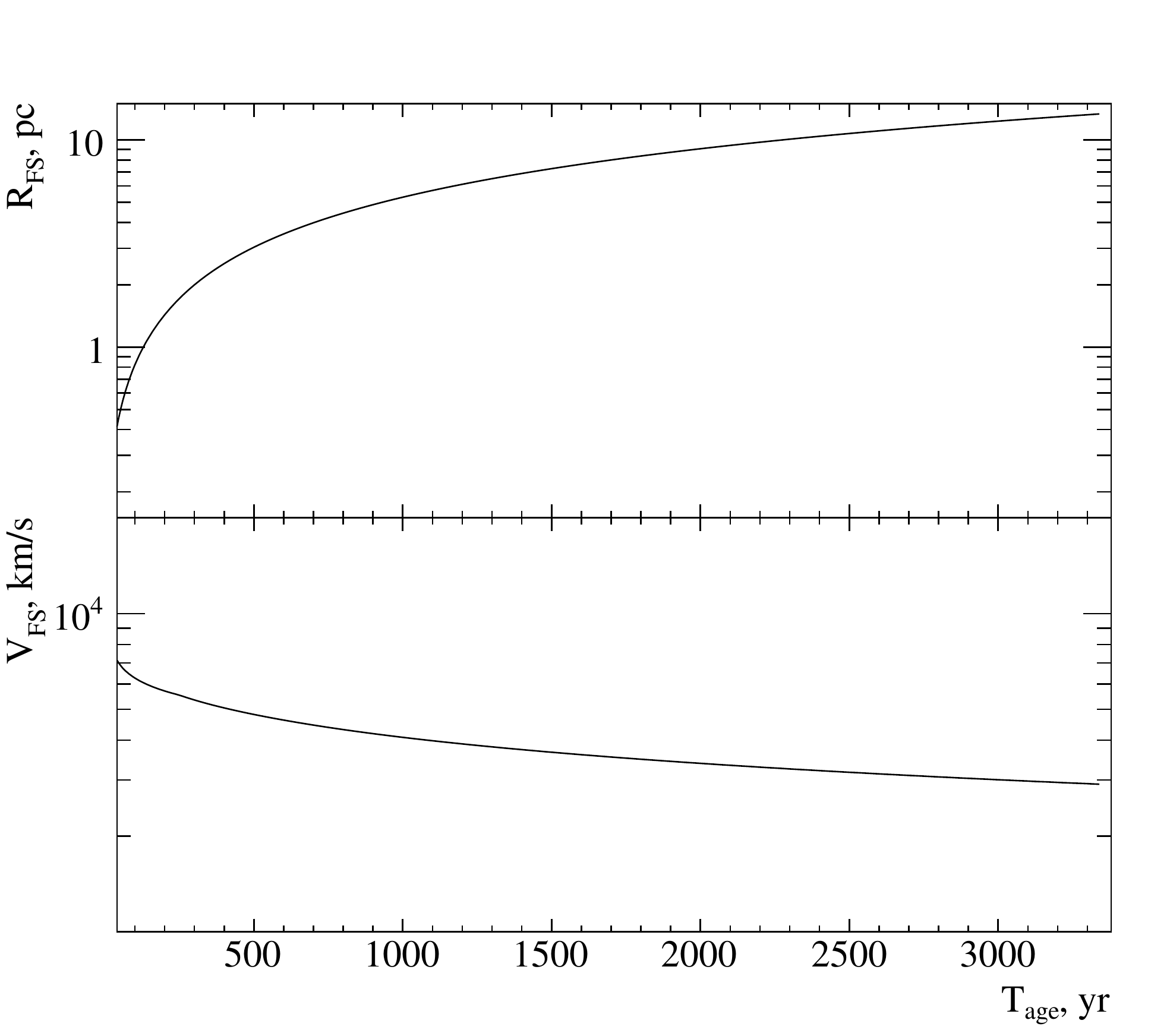}}
  \caption{Time evolution of the forward shock radius (top) and shock velocity (bottom) up to the current age as obtained from our hydrodynamic model.}
  \label{hd_prof}
\end{figure}

\isr{Table \,\ref{hydro} summarizes the properties of the progenitor used as input parameters to our hydrodynamic model and \mpo{lists the} current properties of the SNR obtained from hydrodynamic simulations. The mass-loss rate of the progenitor star, $\dot{M_\star}$, is a free parameter, which is tuned to get the right combination of the current wind density and expansion rate of the remnant. The choice of $\dot{M_\star}$ is, however, in agreement with \mpo{that expected for red supergiants, and it leads to a total mass in the wind of about $13{\rm M}_\odot$ and a time of about 900 kyr to build the wind zone}.} The time evolution of the shock radius and shock velocity is shown on Fig.\,\ref{hd_prof}.

\begin{table}

  \centering
  \caption{Parameters of the hydrodynamic model}
  \begin{tabular}{lc}
    \hline
    \hline
    Parameter & Value\\
    \hline
	$E_{\rm ej}$ [erg] & $10^{51}$ \\	
    $M_{\rm ej}$ [$M_\odot$] & 3 \\
    $v_{\rm wind}$ [km/s]&  15 \\
    $\dot{M_\star}$ [$M_\odot/$yr] & $1.5\times10^{-5}$ \\
    $R_{\rm sh}$ [pc] & 13.3 \\
    $V_{\rm sh}$ [km/s] & 2962 \\
    $n_0$ [cm$^{-3}$] & 0.013 \\
    $T$ [yr] & 3270 \\
    $D$ [pc] & 750\\
    \hline
  \end{tabular}
  \label{hydro}
\end{table}

\subsection{Magnetic field}

\isr{We assume that the magnetic field in the wind zone of the progenitor star, $B_{\rm wind}$, follows a $1/r$ dependence on the radial distance. It should be noted, that the fit of the observational data for \isrr{that} radial dependence of the circumstellar magnetic field results in similar values of the current magnetic field strength as compared to a \mpo{magnetic field of constant strength}. However, stronger synchrotron cooling at earlier stages of the SNR evolution might significantly modify the resulting particle spectrum in the case of the $1/r$ profile.}

We \isr{also} assume that the \isr{circumstellar} magnetic field, \isr{$B_{\rm wind}$}, is amplified upstream of the shock by streaming instabilities. The magnetic field strength in the immediate upstream region is then given by \isr{$B_{\rm u} = kB_{\rm wind}$}. Then for the
shock compression ratio of 4 the immediate downstream magnetic field
strength is \isr{$B_{\rm d} = \sqrt{11} B_{\rm u} = \sqrt{11} k B_{\rm wind}$}. \isrr{It is assumed here that the magnetic field upstream of the shock is turbulent and fully isotropised. In addition to the magnetic turbulence created by streaming instabilities we expect pre-existing turbulence in the stellar wind \citep[see][for the review of the turbulence in the solar wind]{2013LRSP...10....2B}.}

We apply two different parametrizations to describe the distribution of the magnetic field downstream of the shock. In the first scenario we assume that turbulently amplified magnetic field is effectively damped downstream of the forward shock of the remnant. The magnetic turbulence created upstream and at the shock by various instabilities is transferred downstream where it is eventually damped due to the lack of the turbulence driving \citep{2005ApJ...626L.101P}. To describe the magnetic field profile downstream of the shock in this case we adopt a simple parametrization \citep{2005ApJ...626L.101P}
\begin{equation}
  \label{eq:Bdamp}
  B(r) = B_0 + (B_{\rm d} - B_0)e^{(r - R_{\rm sh})/l_{\rm d}},\,r<R_{\rm sh}
\end{equation}
where $B_0$ is the magnetic field far downstream of the shock \isr{and is set in our simulations to be equal to $B_{\rm wind}$} and $l_{\rm d}$ is the damping length scale. \isr{The value of the damping length scale is chosen to be $l_{\rm d}=0.05R_{\rm sh}$ ($\sim0.6$~pc) to fit the peak of the X-ray brightness profile. This large damping scale is consistent with the low magnetic field strength in \velajr\ with correspondingly large eddy size. The damping length can be estimated based on the eddy turn-over time at the largest scales:
\begin{equation}
l_{\rm d}=\chi\,v_\mathrm{d}\,\frac{\lambda}{\delta v}\ ,
\end{equation}
where $\chi$ is a nonlinear efficiency factor on the order of a few, and $v_\mathrm{d}$ is the downstream flow speed.
We can identify the wavelength, $\lambda$, with the Larmor radius of cosmic rays at the highest energy attained in the remnant, $\lambda\simeq r_\mathrm{L} (E_\mathrm{max})$. For the velocity fluctuations we assume energy equipartition between magnetic and kinetic fluctuations,
\begin{equation}
\delta v\simeq \frac{\delta B}{\sqrt{4\pi\rho_\mathrm{d}}}\ .
\end{equation}
Combining all expressions, we find for the damping scale
\begin{equation}
l_{\rm d}\simeq (0.21\ \mathrm{pc})\ \chi\,\frac{E_\mathrm{100}\,n_\mathrm{-1}^{1/2}}{B_\mathrm{-5}^2}\,\left(\frac{v_\mathrm{sh}}{3000\ \mathrm{km/s}}\right)\ ,
\end{equation}
where $E_\mathrm{100}=E_\mathrm{max}/(100 \ \mathrm{TeV})$, $n_\mathrm{-1}=n_\mathrm{-d}/(0.1\ \mathrm{cm}^{-3})$, and $B_\mathrm{-5}=B_\mathrm{d}/(10\ \mu\mathrm{G})$. Noting that all quantities here are measured in the downstream region, except for the shock speed, $v_\mathrm{sh}$, we find with the best-fit parameter values of our model that $l_{\rm d}\simeq 0.6$~pc for a reasonable $\chi\simeq7$.}

Alternatively, we assume that the immediate downstream magnetic field $B_{\rm d}$ is transported inside the SNR with the plasma flow and evolved following the induction equation for ideal MHD.

\begin{figure}[ht!]
  \centering
  \resizebox{\hsize}{!}{\includegraphics{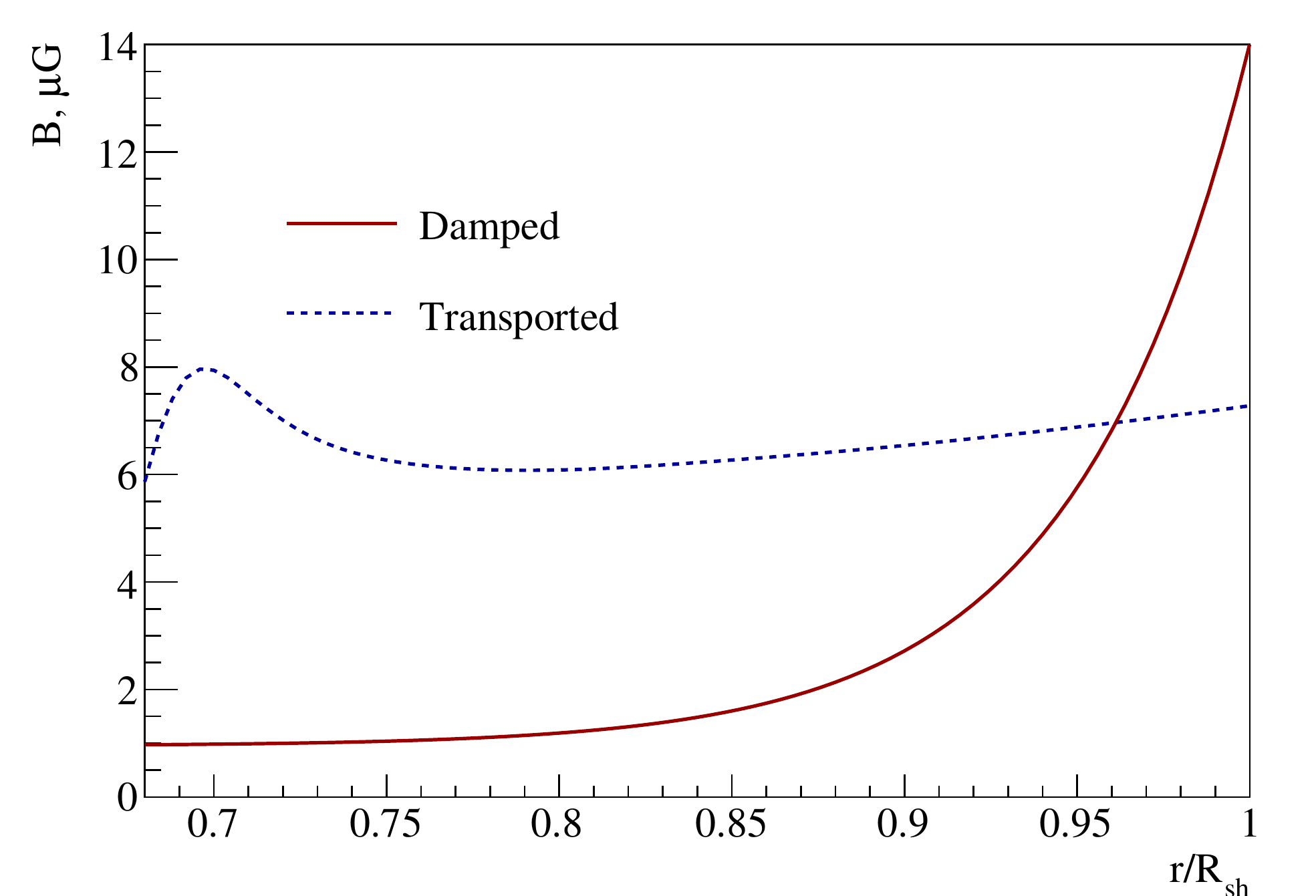}}
  \caption{\isrr{Magnetic field strength as a function of the normalized radius between the contact discontinuity and the forward shock.} Red solid line shows the profile for the magnetic damping scenario and the blue dashed line for the transported magnetic field. Both distributions are shown for the models with the Bohm diffusion coefficient.}
  \label{mf_profile}
\end{figure}

Parameters which describe the magnetic field in both scenarios are collected in the Table\,\ref{models}. Figure\,\ref{mf_profile} illustrates magnetic field profiles in both cases. \isrr{The parameters are chosen in order to fit the broadband} photon spectrum (see Section\,\ref{sec:results})\isr{.}

\isrr{The strength of the magnetic field in the wind at the current stage of evolution adopted in our models is about $1~\mu$G (see Table \ref{models}). This value is in agreement with the experimental measurements of \rob{the} surface magnetic field of red supergiants of $1-10$~G \citep[see e.g.][]{2017A&A...603A.129T}. The assumed $1/r$ profile yields \rob{a} surface magnetic field of $B_\star \simeq (6\ {\mathrm G}) \left({R_\star/(100\,R_\odot)}\right)^{-1}$.}

\subsection{Particle acceleration}

We simulate the particle density evolution solving the transport equation in the form 
\begin{equation}
\frac{\partial N}{\partial t}=\nabla(D\nabla N-\vec{v}N)-\frac{\partial}{\partial p}\left((N\dot{p})-\frac{\nabla \vec{v}}{3}Np\right)+Q,
\label{Transport}
\end{equation}
where $N$ is the differential number density of cosmic rays, $D$ is the spatial diffusion coefficient, $\vec{v}$ is the plasma velocity, $\dot p$ represents the energy losses (in our case synchrotron losses) and $Q$ is the source term.

We use the general form of the spatial diffusion coefficient
\begin{equation}
  D(E) = \zeta D_{\rm Bohm}(E) \left(\frac{E}{1\,{\rm TeV}}\right)^{\alpha} \left(\frac{B}{10\,\mu{\rm G}}\right)^{-\alpha} 
\end{equation}
and consider two different cases: Bohm diffusion with $\zeta = 1$ and $\alpha = 0$ and Kolmogorov-type diffusion with $\alpha = -2/3$ and $\zeta$ chosen in a way to fit the broadband photon spectrum. The type of the diffusion determines the shape of the cut off in the resulting particle spectrum: Kolmogorov-type diffusion implies a slower cut-off than Bohm diffusion. We also calculate the diffusion coefficient by solving the transport equation for magnetic turbulence \isr{dominated} by Alfv\'en waves \citep{2016A&A...593A..20B}. The equation is solved in 1D for spherical symmetry assuming that Alfv\'enic turbulence is isotropic and accounting for compression, advection, cascading, damping and growth due to resonant amplification of Alfv\'en waves. The shape of the cut off of the resulting spectrum of accelerated particles falls in between Kolmogorov-type and Bohm diffusion. In reality, Alfv\'en waves are not the only waves \isr{in} the magnetic turbulence, which might also modify the resulting spectrum. Therefore, two parametrizations considered for the diffusion coefficient can be considered as extreme cases defining the range of possible scenarios.

Particle injection is determined by the source term which is given by
\begin{equation}
Q = \eta n_{\mathrm u}  (V_{\mathrm sh} - \isr{V_{\mathrm wind}}) \delta(R-R_{\mathrm sh}) \delta(p - p_{\mathrm{inj}}),
\end{equation}
where $\eta$ is the injection efficiency parameter, $n_{\mathrm u}$ is the plasma number density in the upstream region, $V_{\mathrm sh}$ is the shock speed, \isr{$V_{\mathrm wind}$ is the wind velocity upstream of the shock}, $R_{\mathrm sh}$ is the shock radius and $p_{\mathrm inj}= \xi p_{\rm th}$ is the injection momentum, defined as a multiple of the momentum \isr{at the} thermal peak of the Maxwellian distribution in the downstream plasma with temperature $T_{\rm d}$,  $p_{\rm th} = \sqrt{2mk_{\rm B}T_{\rm d}}$. We assume the thermal leakage injection model \citep{2005MNRAS.361..907B}, where $p_{\rm inj}$ is the minimum momentum for which a thermal particle can cross the shock and enter the acceleration process and the injection efficiency for the compression ratio of 4 is determined as
\begin{equation}
  \eta = \frac{4}{\sqrt{\pi}}\frac{\xi^3}{e^{\xi^2}}.
\end{equation}
In our simulations we inject all the particles at the same temperature at the position of the shock with the momentum $p_{\mathrm inj}$ assuming that the injection parameter $\xi$ is the same for electrons and for protons. In this case the resulting electron to proton ratio at high energies is determined by the mass ratio, $K_{\rm ep} \simeq \sqrt{m_{\rm e}/m_{\rm p}}$ \citep{1993A&A...270...91P}. The injection parameter $\xi$ is a free parameter which is determined by the fit of the expected emission to the observed gamma-ray data.

We solve the transport equation in 1D in the test particle regime separately for electrons and protons using the RATPaC code as described in \citet{2012APh....35..300T, 2013A&A...552A.102T} taking into account only a forward shock and ignoring a reverse shock as its contribution to the overall particle spectrum is negligible due to the low density in the upstream of the reverse shock \isr{at the current age}. The resulting particle spectrum at the current age of the remnant is then used to simulate non-thermal emission. \isrr{The particle spectrum at low energies follows a power law with a spectral index of 2 because of the test-particle approximation. The choice of the test-particle approximation is well justified as the ratio of the cosmic ray pressure to the ram pressure of the incoming flow, $P_{\rm cr}/P_{\rm ram}$, does not exceed $2\%$ in all our models (see Tables \ref{models} and \ref{anotherhydro})}

\subsection{Non-thermal emission}

We calculate non-thermal emission from the remnant considering three emission processes: synchrotron radiation, IC scattering of accelerated electrons on the Cosmic Microwave Background (CMB) and proton interactions with subsequent pion decay. For IC scattering we do not consider other possible radiation fields as the \velajr\ SNR is located far away from the Galactic center, and local optical and infrared radiation fields are weak and should not play a strong role. Synchrotron emission is calculated accounting for the turbulent magnetic field \citep{2015A&A...574A..43P}.

\begin{table*}

  \centering
  \caption{Models description}
  \begin{tabular}{l|c|c|c|c}
    \hline
    \hline
    &&&\\
    Parameter & BOHM\_BDAMP & BOHM\_BTRAN & KOLM\_BDAMP &KOLM\_BTRAN\\
    
    \hline
    $\alpha$ & 0 & 0& $-2/3$ & $-2/3$ \\  
    $\zeta$& 1& 1& $25.0$ & $22.2$\\ 
    $\xi$& 4.23 & 4.20 & 4.14 & 4.12 \\
    $K_{\rm ep}$ & $0.022$& $0.021$& $0.025$& $0.026$ \\
    Magnetic field  & damped  & transported & damped & transported\\
    $B_{\rm wind},\,\mu$G & 1.0 & 1.3 & 1.2 & 0.9 \\
    $k$ & 4.4 & 1.7 & 2.9 & 2.2\\
    $B_{\rm d},\,\mu$G & 14.0 & 7.3 & 11.7 & 6.5 \\
    $B_0,\,\mu$G & 1.0& --- & 1.2 & --- \\
    $l_{\rm d}$ & 0.05& ---  & 0.05 & ---\\ 
    \hline
    $P_{\rm cr}/P_{\rm ram}$ & 0.013&  0.014& 0.015& 0.016\\
    $W_{\rm tot,pr}$, erg & $1.2 \times 10^{49}$& $1.4 \times 10^{49}$& $1.6 \times 10^{49}$& $1.8 \times 10^{49}$\\
    $W_{\rm tot,el}$, erg & $4.1 \times 10^{47}$& $5.4 \times 10^{47}$& $7.1 \times 10^{47}$ & $8.0 \times 10^{47}$\\
    \hline
  \end{tabular}
  \label{models}
\end{table*}

\section{Results}
\label{sec:results}

\is{In this section we present the results of our modeling of the broadband emission from the \velajr\ SNR. The emission from the remnant is calculated assuming spherical symmetry. We construct four models based on our assumptions of the form of the spatial diffusion coefficient and the shape of the magnetic field distribution downstream of the shock. Main parameters used in these models are collected in the Table\,\ref{models}.}

\subsection{Broadband photon spectrum}

\begin{figure*}[ht!]
  \centering
  \centering
  \resizebox{\hsize}{!}{\includegraphics{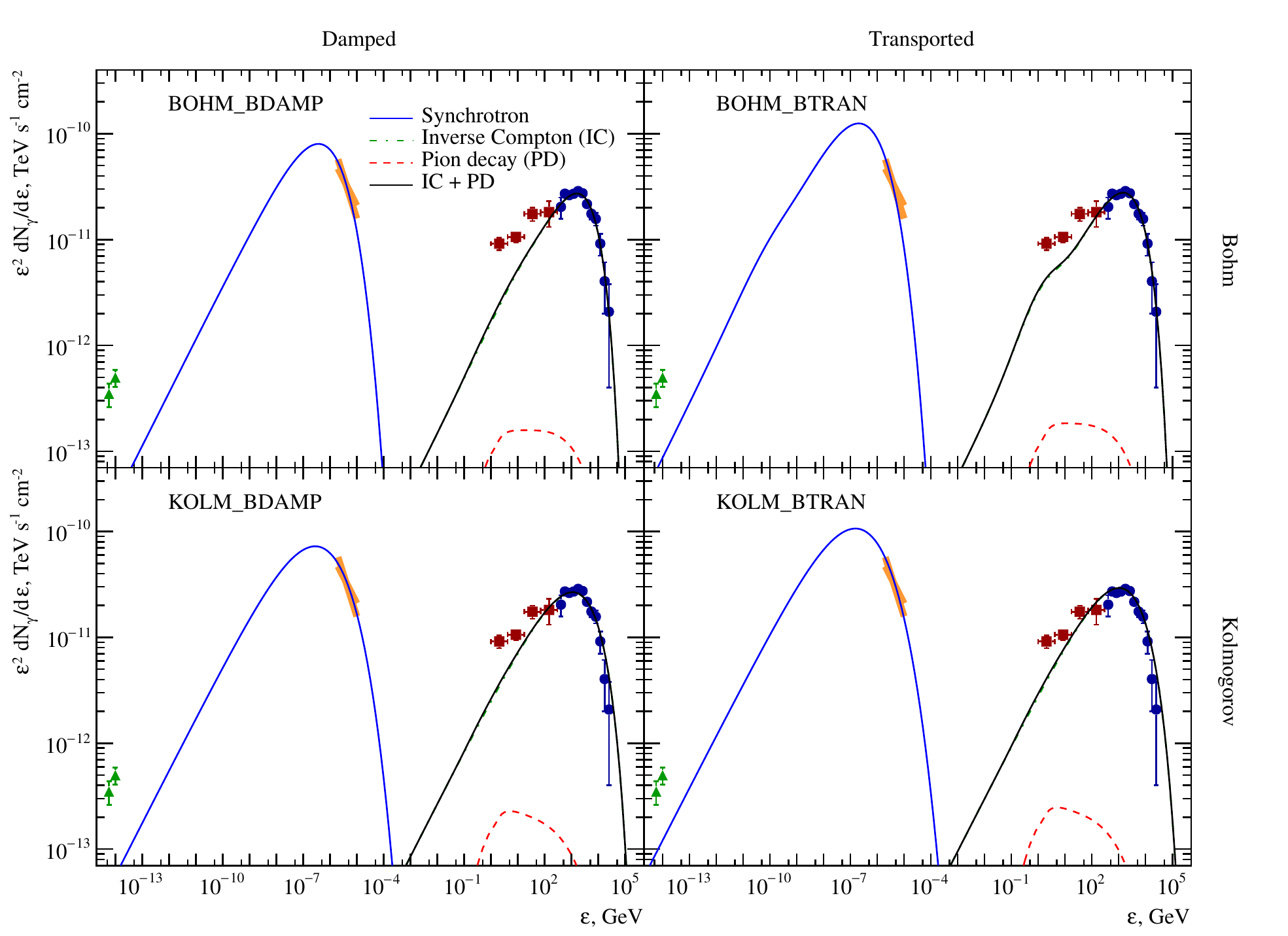}}
  \caption{Spectral enery distribution of the \velajr\ SNR. Lines represent simulated emission from the source produced in different processes as specified in the legend in the magnetic field damping model (left panels) and synchrotron cooling model (right panels). Green triangles indicate the observed radio emission as detected by Parkes \citep{paper:vjr_radio_points}, an orange bow-tie shows the spectral fit of the X-ray \xmm\ data \citep{paper:vjr_hess_paper2}, red squares represents the \fermi\ data points \citep{paper:vjr_fermi} and blue circles show the \hess\ data points \citep{2016arXiv161101863H}.}
  \label{sed}
\end{figure*}

Figure\,\ref{sed} shows the observed broadband spectral energy distribution (SED) overlaid with the curves representing simulated emission in four different models. In all four models the simulated SED strongly underpredicts the observed
radio emission. This discrepancy is further discussed in Section\,\ref{rademission}.

\begin{figure}[ht!]
  \centering
  \resizebox{\hsize}{!}{\includegraphics{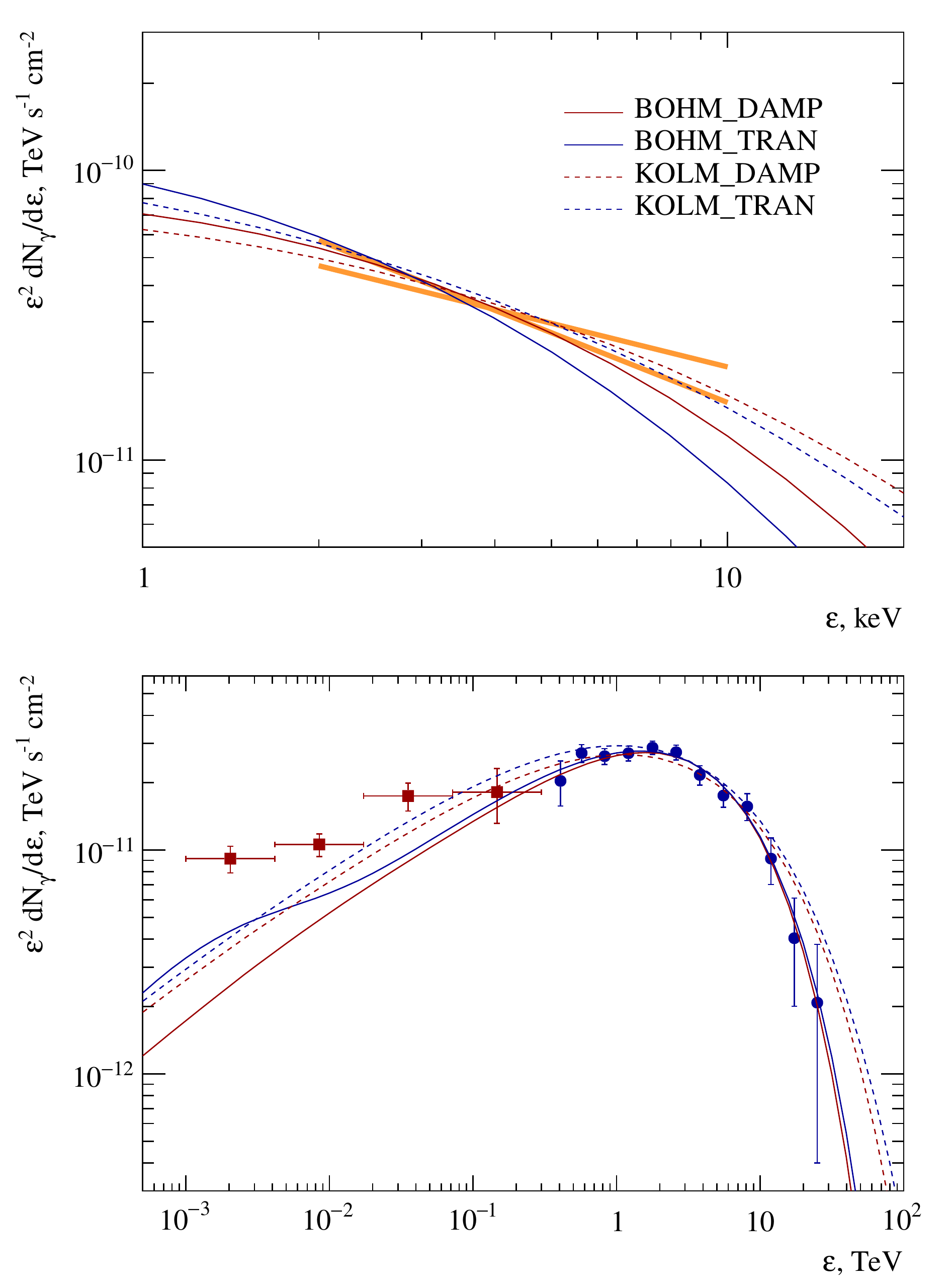}}
  \caption{Zoom into X-ray (top panel) and GeV-TeV (bottom panel) spectra for all four models.}
  \label{sed_zoom}
\end{figure}

The choice of the spatial diffusion coefficient determines the shape of the cutoff in the electron spectrum and thus the shape of the X-ray and gamma-ray spectra. In the case of the Bohm diffusion we fail to reproduce the slope of the X-ray spectrum as the synchrotron emission exhibits a faster than observed cutoff of the spectrum (Fig.\,\ref{sed_zoom}, top panel). Our BOHM models also slightly underpredict GeV emission but at the same time provide a very good fit of
the TeV spectrum (Fig.\,\ref{sed_zoom}, bottom panel). On the other hand, the models with the Kolmogorov-type diffusion coefficient much better reproduce the observed X-ray slope and GeV data, but yield a slightly worse fit of the cutoff at VHEs, although still compatible with data. The observational data suggest a faster cutoff in the gamma-ray spectrum with respect to the X-ray spectrum. In addition, the detection of hard X-rays from the remnant with \suzaku\ \citep{2016PASJ...68S..10T} also suggests a much slower cutoff in the X-ray spectrum which is much better reproduced by the Kolmogorov-type diffusion coefficient. The observed GeV data suggest a softer electron spectrum than the one obtained in DSA in the test-particle approximation. However, it is still possible to reproduce the observed GeV spectrum with the Kolmogorov-type diffusion coefficient due to the lower cutoff energy and slower cutoff.  

\isr{The IC emission in the BOHM\_BTRAN model \isrr{shows a shoulder around 1 GeV due to synchrotron cooling.} This feature is created because of the $1/r$ profile of the ambient magnetic field. At the early stages of SNR evolution the ambient magnetic field is very strong, \isrr{cooling electrons efficiently}. This feature is not that prominent in the synchrotron spectrum because the cooled particles responsible for this break are accumulated at the contact discontinuity where the magnetic field is lower and hence the synchrotron emission is weaker. It is also not visible in KOLM\_BTRAN model due to a weaker ambient magnetic field. The impact of the ambient magnetic field on the broadband radiation spectrum will be further discussed in a follow-up paper.}

The contribution of the pion decay process to the overall gamma-ray flux is negligible for the type of injection we implement. If we relax the requirement that injection parameter is the same for electrons and protons, and reduce the electron-to-proton ratio the observed SED could be reproduced with a higher contribution from the pion decay in the gamma-ray energy range. However, it is easy to see that to match the level of the \isr{GeV-TeV} emission exclusively by the pion decay process would require a total energy in protons to be comparable to or higher than the supernova explosion energy as this would require $\sim100$ times more protons.

The difference between the models with damped and transported magnetic fields is not that pronounced in the total volume integrated spectral distribution of the SNR. However, as it is shown in the next subsection, significant differences  arise in the morphology of the remnant and spatial spectral variations. It should be noted though, that the fit of the observed SED requires different values of the magnetic field strength downstream of the forward shock in these two scenarios (see Table\,\ref{models}). Damping models require a stronger field in the immediate downstream region. 

Our results are sensitive to variations of the model parameters. Although small changes do not strongly modify the models, the parameter space is well constrained by the observational data. The X-ray to gamma-ray flux ratio constrains the magnetic field strength which in the immediate downstream cannot be higher than $\sim 15\,\mu$G even in the damped field scenario. Magnetic field strength in turn determines \isrr{the total energy in accelerated electrons}. We also notice that the maximum energy to which particles can be accelerated is sensitive to the magnetic field strength in the far upstream of the shock as it determines the diffusion of particles away from the shock. The sensitivity of our results on the parameters of the hydrodynamic model is discussed in Section\,\ref{anotherhydro}.

\subsection{X-ray morphology and spatial spectral \isr{variations}}

\isr{To study the morphology of the remnant we produce 2D energy dependent intensity maps with a cell size of $0.67^\prime$, equal to the width of the regions $1$ to $4$ as defined for the X-ray spectral analysis in \citet{2013A&A...551A.132K}. Then for the cells \mpo{falling into these} regions we extract the flux and the spectrum in the energy range between 2 keV and 10 keV. The regions 5 to 9 have twice larger width, and for these regions we combine the emission from two cells for the spectrum estimation.}

\begin{figure}[t!]
  \centering
  \resizebox{\hsize}{!}{\includegraphics{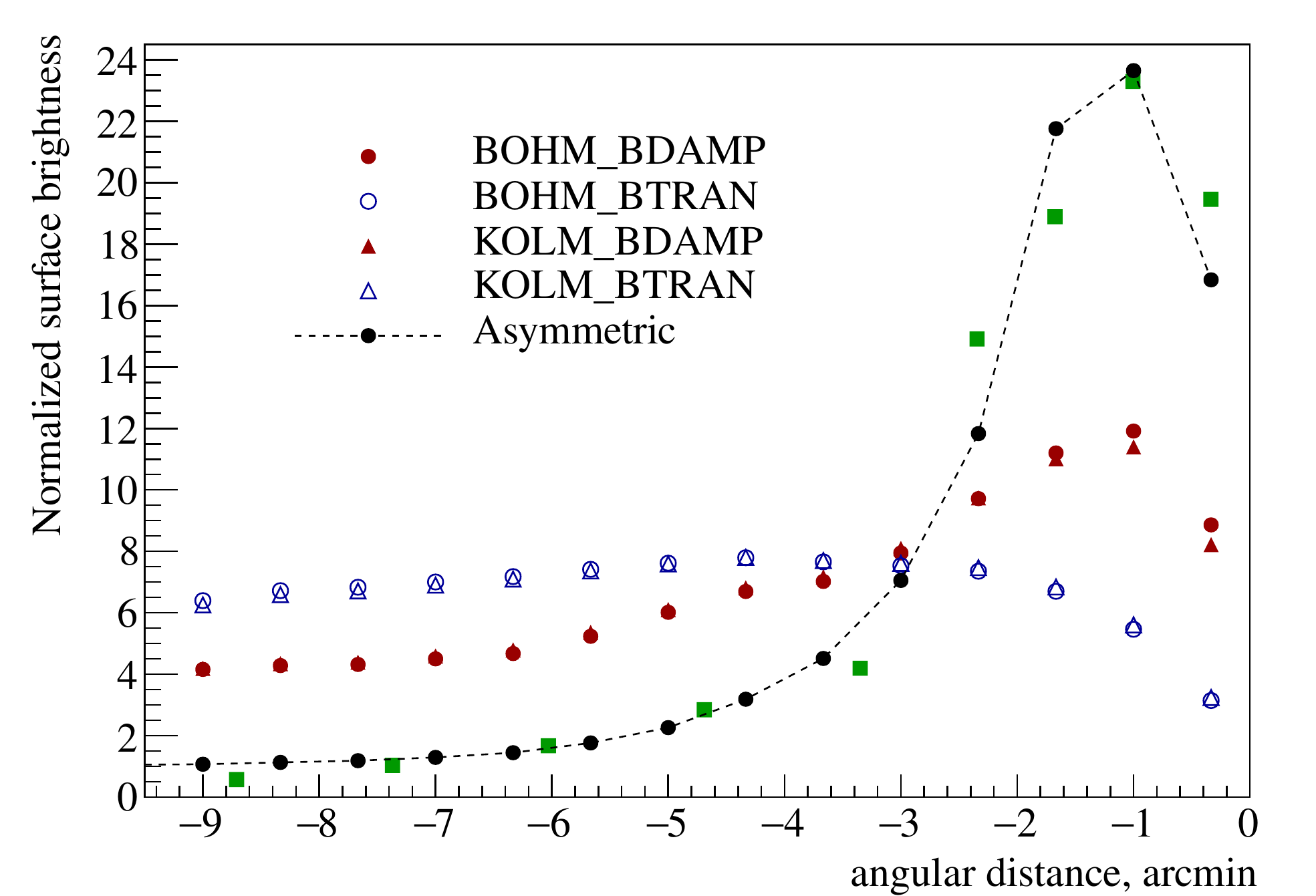}}
    \caption{Simulated \isr{projected} X-ray radial flux profiles for different models \isr{assuming spherical symmetry} compared to the experimental measurements \citep[green squares][]{2013A&A...551A.132K}. Normalized \isrr{surface brightness} in the energy range $2-10$\,keV is plotted \isrr{as} a function of \isrr{the angular distance from the forward shock of the SNR with negative values corresponding to the interior of the remnant}. \isr{The radial profile obtained within the asymmetric model discussed in Section\,\ref{asymmetry} is represented by the black filled circles connected with the dashed line.}}
  \label{flux_profile}
\end{figure}

\isr{Figure\,\ref{flux_profile} \isrr{compares} the simulated surface brightness profile to the observed one. For a direct comparison the observed flux \citep{2013A&A...551A.132K} was transformed into the surface brightness accounting for the different size of the regions.} All profiles are normalized in a way that the integral along the radius of the remnant above $0.85\,R_{\rm sh}$ \isr{(above $51^\prime$ for the radius of $1^\circ$)} is equal to unity. \isr{The difference between Bohm and Kolmogorov-type models is almost negligible, but the shape of the brightness profile strongly depends on the magnetic field model. The models with the magnetic damping scenario (red markers) better reproduce the shape of the profile with the emission peaking around $1^\prime$ away from the shock.} In the case of the transported magnetic field \isr{(blue markers)} the peak is located closer to the center of the remnant at \isr{the angular distance of} around $(3^\prime - 5^\prime)$ from the shock. \isrr{Neither case, however, can reproduce the magnitude of the flux decrease.} \isr{While in the case of the transported field scenario the magnetic field stays almost constant downstream of the shock \isrr{resulting} in substantial synchrotron emission from the interior, in the damped field scenario the magnetic field strength decreases very quickly with the distance from the shock, suppressing synchrotron emission, and hence the plateau in the profile is mainly determined by the projection effect \mpo{and therefore inevitable under spherical symmetry}. Therefore, the observed decrease of the X-ray intensity towards the center of the remnant could indicate a strong deviation from spherical symmetry. This point is further discussed in the Section\,\ref{asymmetry}.}

\isr{To study the spectral variations with the distance from the shock we extract the spectrum for each region and fit it in the same way as \isrr{\citet{2013A&A...551A.132K}.}} Simulated spatial distributions of the spectral index and of the cut-off energy in all four models 
follow the same trend as the observed radial distributions. However the observed variation of the spectral parameters is stronger. Close to the shock the spectral variation is stronger in the case of the damped magnetic field profile, but at the distance $\gtrsim3^{\prime}$ the spectral shape stays roughly constant while in the case of the transported field the spectrum keeps getting softer with the distance from the shock. This happens because in the case of the damped profile the magnetic field strength decreases much faster with distance from the shock and already at $\sim3^\prime$ the spectrum is fully determined by the projection effect. In the case of the transported field the magnetic field strength drops more slowly and the interior of the remnant still significantly contributes to the integrated intensity along the line of sight.

\begin{figure}[ht!]
  \centering
  \resizebox{\hsize}{!}{\includegraphics{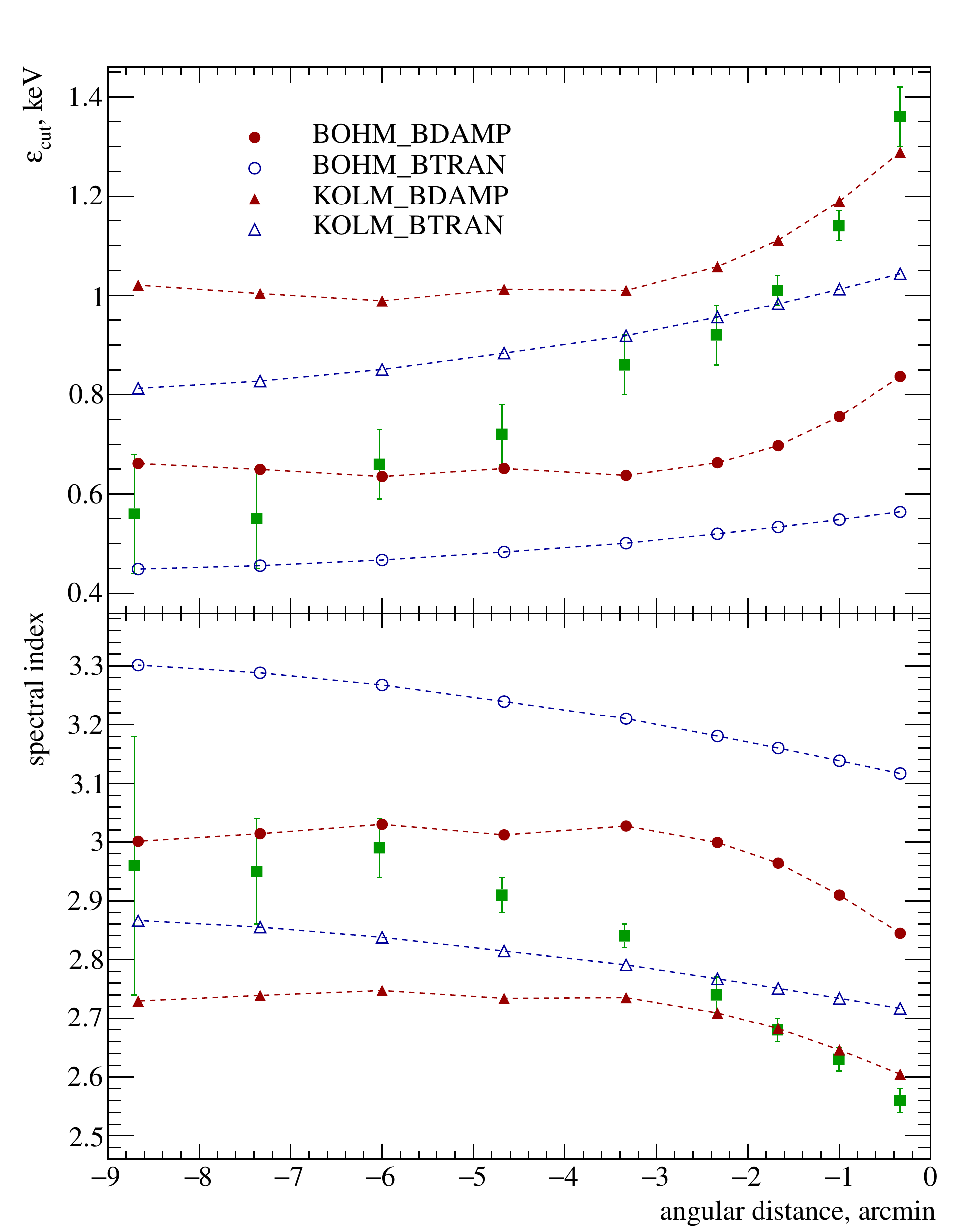}}
    \caption{Simulated \isr{projected} radial profiles of the X-ray spectral parameters for four different models \isr{assuming spherical symmetry}. The top panel shows the cut-off energy and the bottom one shows the spectral index \isrr{in the $2-10$ keV band} as a function of the angular distance from the forward shock of the SNR. \isrr{The distributions are obtained by extracting the spectrum for each region and fitting it with an exponentially cut-off power law assuming the photon spectral index of $\Gamma=1.6$ (top panel) and with a simple power law (bottom panel).} In both panels green squares indicate the observational data \citep{2013A&A...551A.132K}.}
  \label{spec_profiles}
\end{figure}

Distributions resulting from models using Bohm and Kolmogorov-type diffusion coefficients basically mimic each other, but differ in the parameter values. Kolmogorov-type diffusion much better reproduces the spectral shape close to the forward shock where most of the X-ray emission is coming from. This is not surprising as the \isr{Kolmogorov-type models fit better} the spectrum of the total X-ray emission, and the total emission is dominated by the emission from the shock region.

\subsection{Gamma-ray morphology}

\begin{figure}[ht!]
  \centering
  \resizebox{\hsize}{!}{\includegraphics{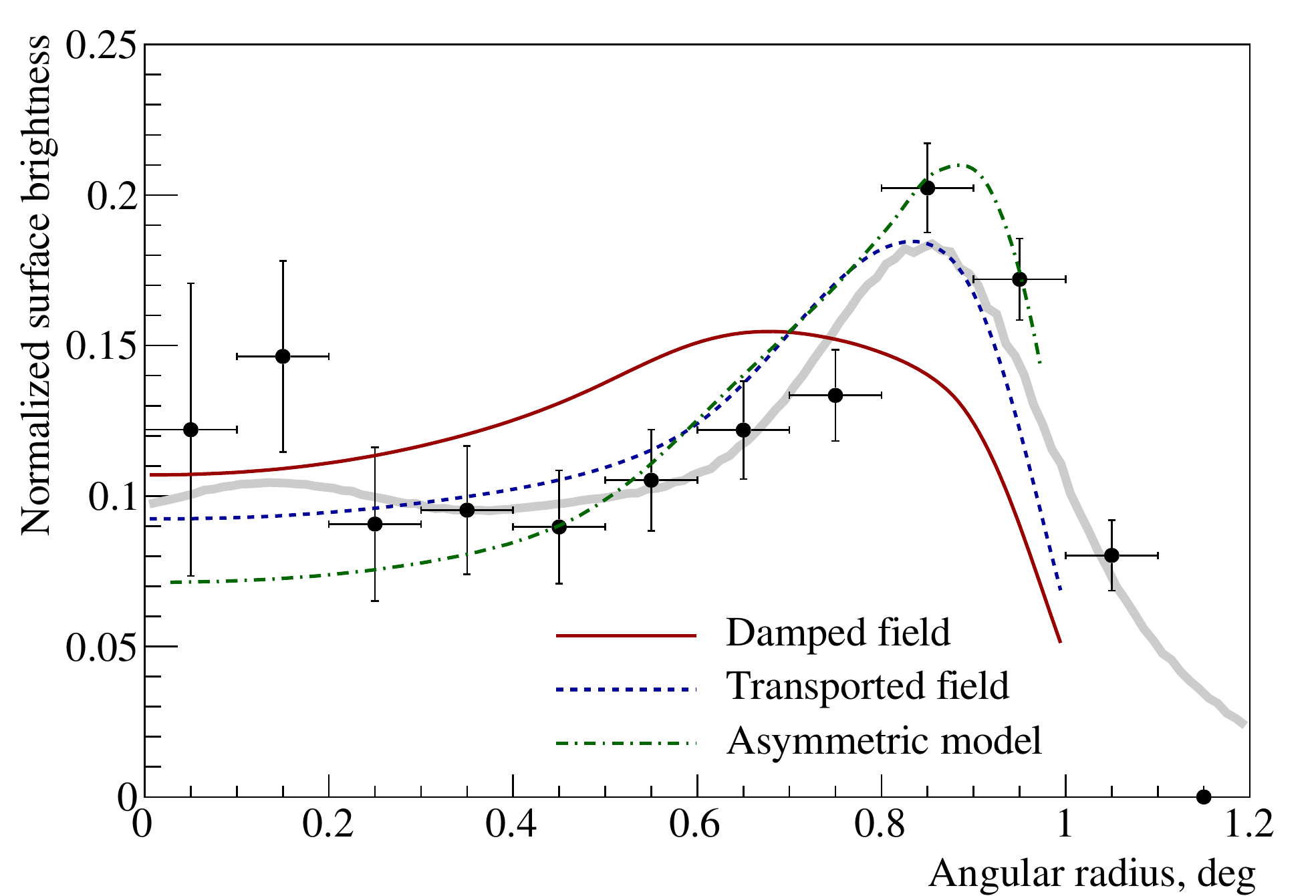}}
  \caption{\isr{Simulated} radial surface brightness profiles of the gamma-ray emission in the 300 GeV $-$ 20 TeV energy range compared to the \hess\ measurements \isr{for the northern part of the remnant} \citep[][black filled circles]{paper:vjr_hess_paper2}. Red solid and blue dashed lines show simulated profiles for the \isr{spherically symmetric} models BOHM\_DAMP and BOHM\_KOLM respectively. The radial profile obtained within the asymmetric model discussed in Section\,\ref{asymmetry} is represented by the green dash-dotted line. \isr{This profile is averaged over the azimuth angle for the hemisphere containing the cone of enhanced emission.} \isr{The thick gray line shows the azimuthally averaged radial profile for the northern part of the SNR extracted from the recently published updated H.E.S.S. excess skymap \citep{2016arXiv161101863H}. This skymap is smoothed with a Gaussian function of width $0.08^\circ$}. All simulated profiles are normalized in the same way as \hess\ measurements \isr{in \citet{paper:vjr_hess_paper2}}.}
  \label{tev_profile}
\end{figure}

In Figure\,\ref{tev_profile} we compare our simulated gamma-ray profiles to the TeV \isr{profiles} measured by H.E.S.S. Simulated profiles in the energy range from 300\,GeV to 20\,TeV are extracted from the
simulated surface brightness map smoothed with a Gaussian kernel with $\sigma = 0.06^\circ$, which roughly resembles the \hess\ PSF \isr{in the \citet{paper:vjr_hess_paper2} analysis}. The \hess\ profile was normalized \isr{so that} the sum of the histogram contents between $0.3^\circ$ and $1.2^{\circ}$ is equal to unity \citep{paper:vjr_hess_paper2}. Simulated profiles were normalized in the same way for the direct comparison. We show the profiles only for the models with Bohm diffusion coefficient (red solid and blue dashed lines; green dash-dotted line represents the asymmetric model discussed in Section\,\ref{asymmetry}) as the ones for the Kolmogorov-type diffusion are very similar.

It is clear that the model with the transported magnetic field resembles the data much better than the magnetic damping model. In the magnetic damping scenario the peak appears to be significantly broader and the
decrease of the emission towards the center of the remnant is slower. This is due to a much weaker synchrotron cooling of the particles downstream of the shock at distances $r/R_{\rm FS} \lesssim 0.9$. This results in a larger number of high energy particles inside the remnant and in turn \rob{a} broader radial distribution.

This result is in conflict with the result for the X-ray morphology which is better reproduced by the magnetic damping scenario. For the synchrotron emission in the magnetic damping scenario the key parameter is the magnetic field strength which drops exponentially with the distance from the shock and thus determines the shape of the radial profile.

\section{Discussion}
\label{discussion}

\subsection{Radio emission}
\label{rademission}
All four models strongly underpredict the radio emission from \velajr\ This can be an indication that our models do not account for some other possible processes which can take place in SNRs. \isr{For example stochastic re-acceleration of particles at the turbulence in the immediate downstream region of the shock \citep{2015A&A...574A..43P} can soften the resulting \isrr{synchrotron} spectrum, }which means that it would be possible to explain the radio emission within the models presented here by simply adding another process.

Another possible approach to reproduce the radio emission from \velajr\ would be to account for the nonlinear feedback of the cosmic-ray pressure on the hydrodynamic evolution of the remnant. In this case cosmic rays would modify the shock and in turn the particle spectrum resulting in a softer spectrum. For example, \citet{2013ApJ...767...20L} succeeded in reproducing the radio emission together with the X-ray emission and gamma-rays using the nonlinear DSA calculations with a semi-analytic solution \citep[e.g.][]{2005MNRAS.361..907B, 2009MNRAS.395..895C}. A drawback of their model is that they do not simulate the shape of the cutoff of the particle spectrum but simply assume it to follow $\propto \exp(-(p/p_{\rm max})^\beta)$, where $p_{\rm max}$ is the maximum momentum of the particle and is obtained from simulations and $\beta$ is simply chosen in such a way that the model explains the X-ray spectrum. Therefore, in this approach the model parameters can be tuned in a way to obtain a right $p_{\rm max}$ to fit the radio points
even for the particle spectral index of 2. 

Another effect which was not considered here and can cause a softening of the particle spectrum is Alfv\'enic drift \citep[see e.g.][]{2012A&A...538A..81M}. However, the magnetic field suggested by the simultaneous fit of the X-ray and gamma-ray emission is too low to cause this effect, because the Alfv\'en velocity is $V_{\rm A}\sim100$\,km/s. For a higher magnetic field strength radio and X-ray spectra could possibly be fit even without Alfv\'enic drift as in this case strong synchrotron cooling would create a break in the synchrotron spectrum. However, stronger magnetic field would require fewer electrons to produce the X-ray emission and thus the model would strongly underpredict the gamma-ray emission.

It should also be noted that the reported value of the detected radio flux can be misleading as the remnant is located in a very crowded region with a very complicated background and there are indications that a part of the radio continuum emission coincident with \velajr\ might be associated with the Vela SNR \citep{maxted_radio}.

\isr{High resolution radio observations and detailed morphological studies could be of great importance for constraining the magnetic field downstream of the shock. Damped and transported magnetic field profiles predict different radio morphology of the remnant. In the case of the damped field, the radio emission from the SNR similarly to the X-ray emission would be determined by the strength of the magnetic field hence the radio morphology of the remnant should be similar to the X-ray morphology. In the transported field scenario the magnetic field strength is almost constant downstream (see Fig. \ref{mf_profile}) hence the morphology is determined mainly by the number of particles emitting synchrotron radiation. The X-ray emission would still peak close to the shock because of the lack of high energy electrons far downstream due to synchrotron cooling, but the radio emission is expected to have its maximum closer to the contact discontinuity due to a high abundance of low energy electrons combined with the projection effect. Radio images obtained with the 64-m Parkes radio telescope \citep{paper:vjr_radio_points} show a good correlation of the radio emission with the X-ray emission, suggesting that the magnetic field should quickly decrease downstream, but more quantitative comparison is difficult due to a poor angular resolution of radio observations and possible contamination by the emission from the Vela SNR.}

\subsection{Deviation from spherical symmetry}
\label{asymmetry}

The observed X-ray flux profile exhibits a very fast decrease of the flux towards the interior of the remnant. This is a strong indication for a significant deviation from spherical symmetry. Indeed, even if we assume that the magnetic field strength falls to zero and that electrons capable of radiating X-ray synchrotron can be found only in the immediate vicinity of the shock, still it would be impossible to obtain such a decrease within the assumption of spherical symmetry simply due to the projection effect. The projection effect also \isr{causes} a slower softening of the spectrum towards the interior as the projected emission from regions close to the shock \isr{dominates the total projected emission}. Therefore, the observed softening of the spectrum which is faster than the one simulated within the spherical symmetry assumption also suggests the asymmetry of the remnant. Finally, the observed X-ray morphology of the remnant also deviates from spherical symmetry showing a number of regions with local brightening of the X-ray emission, one of which was used by \citet{2013A&A...551A.132K} to detect the spectral softening.

To test the asymmetry hypothesis and check how well we can reproduce the morphology we assume that in one radial direction within the solid angle $\Omega$ the synchrotron emission is brighter than in the rest of the remnant. This could be achieved, for example, if in this cone the injection of particles was more efficient due to higher ambient density in this region. In this case the radial profile taken in this direction would much better resemble the data as the contribution of the shock region emission to the projected emission inside the remnant would be lower. We solve the transport equation in 1D assuming there is no tangential transport of particles and therefore we can simply modify one of our models calculated for spherical symmetry by assuming that synchrotron emission is stronger in the region limited by the cone with the solid angle $\Omega$. For this exercise we use model KOLM\_BDAMP as it reproduces both the right position of the peak in the X-ray flux profile and the X-ray spectral index value in the immediate downstream of the shock. We assume that the synchrotron emission within the solid angle $\Omega = 2\pi(1 - \cos{\theta})$ is stronger than in the rest of the remnant by a factor $\kappa$, but at the same time keep the total synchrotron emission from the remnant constant, so that our modified model would still reproduce the SED in the same way as KOLM\_BDAMP. \isrr{The opening angle $\theta$ is set by the size of the bright filament, which roughly corresponds} to the size of the field of view studied in \citet{2013A&A...551A.132K}.

\isr{The simulated radial profile of the filament for $\theta = 15^\circ$ and $\kappa = 12$ is shown with black filled circles connected with the dashed line on Fig.\ref{flux_profile} compared to the observed radial profile. The regions are defined in the same way as for the sperically symmetric models.} 
It can be seen that the modified asymmetric model can reproduce the observed radial profile fairly well reproducing both the shape of the profile and the magnitude of the flux decrease.

\begin{figure}[ht!]
  \centering
  \resizebox{\hsize}{!}{\includegraphics{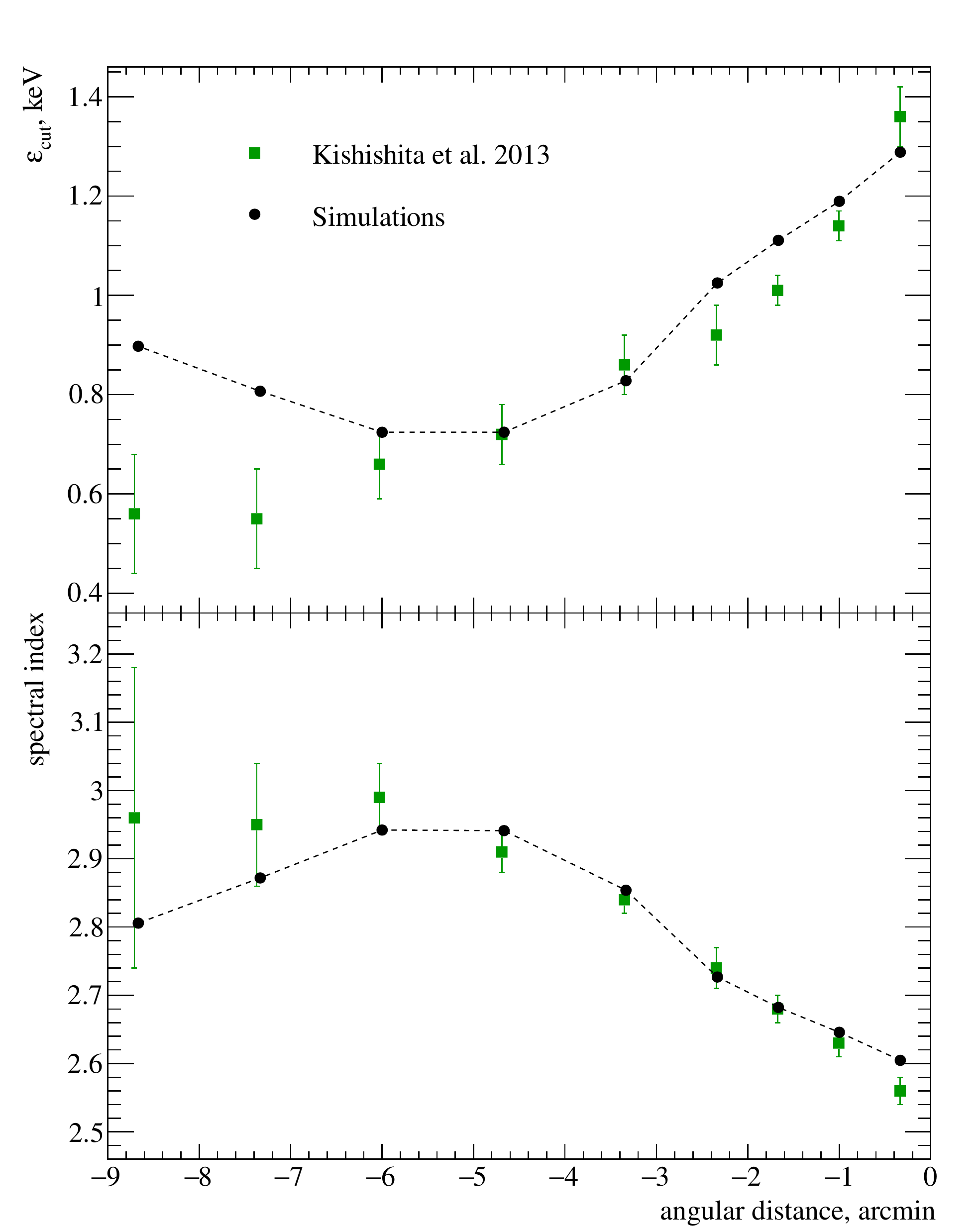}}
  \caption{Simulated cut-off energy (top panel) and spectral index (bottom panel) \is{of the X-ray spectrum in the filament as a function of the angular distance from the shock} compared to the experimental measurements. \is{Simulated distributions are obtained using the \isr{asymmetric model described in \isrr{Section~\ref{asymmetry}}.}}}
  \label{cone_spec_profile}
\end{figure}

As expected, by improving the fit of the \isr{surface brightness} profile, we can also better reproduce the observed softening of the spectrum. Figure\,\ref{cone_spec_profile} shows simulated radial distributions of the spectral index and the cut-off energy compared to the observational data. Simulated distributions start to deviate from the measured values at $\gtrsim5^{\prime}$ from the forward shock. This happens because at this distance the contribution from the cone of the brighter emission becomes negligible in comparison to the rest of the remnant \isr{along} the line of sight and thus simulated spectral parameters start to resemble the spherically symmetric model. This results in a upturn for the cut-off energy and downturn for the spectral index at $\sim5^{\prime}$ which is not seen in spherically symmetric model because the emission is evenly distributed.

If the brightening of the synchrotron emission is due to the higher number of electrons in the region limited by the cone, this would also naturally enhance the inverse Compton emission from that region by the same factor\isrr{. Observationally, this} would result in a local brightening of the gamma-ray emission from the location of the X-ray filament which is in agreement with the observed correlation between the TeV and X-ray morphology. Remarkably, the choice of the cone parameters (the opening angle $\theta$ and the brightening factor $\kappa$), motivated by the size and the properties of the X-ray filament, results also in a much better fit of the observed azimuthally averaged surface brightness profile of the gamma-ray emission above $300$\,GeV as compared to the model with the damped magnetic field under the spherical symmetry assumption (Fig.\,\ref{tev_profile}). \is{This means that for the asymmetric case the damped field model can fit both X-ray and gamma-ray profiles as opposed to the spherically symmetric case where the gamma-ray intensity profile prefers the transported field model.}

\isr{Our cone-model for the X-ray filament appears to also be in agreement with the azimuthal intensity distribution of the X-ray and TeV emission. \isrr{\citet{2016arXiv161101863H} provide azimuthal intensity profiles} for an annulus with an inner radius of $0.6^\circ$ and outer radius of $1^\circ$ around the center of the SNR split in forty equisized bins. Defining the bins in the same way we extract the flux from the bin containing the bright filament and compare it to the flux from the bin outside the cone of the brighter emission. The azimuthal intensity contrast estimated as the ratio of these two fluxes is about five (for both X-ray and TeV emission) which is in agreement with the observed ratio.}

As mentioned above, \citet{2017arXiv170807911F} detected a good spatial correlation of the TeV emission with the distribution of the HI gas indirectly suggesting the hadronic scenario for the gamma-ray emission because of the higher density of the ambient medium. This, however, also implies more efficient particle injection and thus might explain local brightening of X-ray and TeV emission within the leptonic scenario simply because of the higher number of accelerated electrons. At the same time, gamma-ray emission produced in hadronic interactions might still be too low to significantly contribute to the observed gamma-ray emission. Therefore the HI gas does not necessarily indicate the hadronic scenario, but it might be the reason \isr{for} the existence of regions with local brightening of the emission. This idea is also supported by the coincidence of the HI emission with a radio filament as noticed by \citet{maxted_radio}.

\subsection{Uncertainties of the hydrodynamic model}
\label{anotherhydro}

To test how sensitive our results are to the parameters of our hydrodynamic model we built an alternative model based on the expansion rate measured using
the \chandra\ data \citep{2015ApJ...798...82A}. We kept the same ejecta mass, $M_{\rm ej} = 3\,M_\odot$, and wind speed, $v_{\rm wind} = 15$\,km/s, and varied all other
parameters to obtain the right angular size of the remnant and to reproduce the observed expansion rate. As before we constrain the ambient density by
the lack of X-ray thermal emission and set the upper limit on the distance at $1$\,kpc. The parameters of the alternative model are listed in Table\,\ref{hydrotwo}. A lower expansion rate naturally yields a lower shock speed and higher age of the remnant. To keep the angular size unchanged we also need to slightly increase the density, the distance to the remnant and the mass-loss rate of the
progenitors wind. \isr{A higher mass-loss rate, however, implies a rather high total mass in the wind of about $33{\rm M}_\odot$.}

\begin{table}

  \centering
  \caption{Parameters of the alternative hydrodynamic model}
  \begin{tabular}{lc}
    \hline
    \hline
    \\
    Parameter & Value\\
    \hline
    $E_{\rm ej}$ [erg] & $10^{51}$ \\	
    $M_{\rm ej}$ [$M_\odot$] & 3 \\
    $v_{\rm wind}$ [km/s]&  15 \\
    $\dot{M_\star}$ [$M_\odot/$yr] & $3.8\times10^{-5}$ \\
    $R_{\rm sh}$ [pc] & 16.2 \\
    $V_{\rm sh}$ [km/s] & 1815 \\
    $n_0$ [cm$^{-3}$] & 0.021 \\
    $T$ [yr] & 6120 \\
    $D$ [pc] & 900\\
    \hline
  \end{tabular}
  \label{hydrotwo}
\end{table}

Based on the new hydrodynamic model we create a modified model of the broadband emission for the Bohm diffusion coefficient and damped magnetic field. The model parameters are listed in the Table\,\ref{modmodel}. We tune the parameters \isrr{so} that the model provides a fit to the data similar to that of the primary model. It appears that the fit requires only small changes of the main parameters (Table\,\ref{modmodel}).

\begin{table}

  \centering
  \caption{Description of the emission model for the alternative hydrodynamic solution}
  \begin{tabular}{lc}
    \hline
    \hline
    \\
    Parameter & Value \\
    
    \hline
    $\alpha$ & 0  \\  
    $\zeta$& 1\\
    $\xi$& 4.23 \\
    $K_{\rm ep}$ & $0.047$\\
    Magnetic field  & damped\\
    $B_{\rm wind},\,\mu$G & 2.0\\
    $k$ & 2.2 \\
    \isr{$B_{\rm d},\,\mu$G} & 14.6  \\
    $B_0,\,\mu$G & 2.0 \\
    $l_{\rm d}$ & 0.05\\ 
    \hline
    $P_{\rm cr}/P_{\rm ram}$ & 0.019\\
    $W_{\rm tot,pr}$, erg & $1.3 \times 10^{49}$\\
    $W_{\rm tot,el}$, erg & $6.6 \times 10^{47}$\\
     \hline
  \end{tabular}
  \label{modmodel}
\end{table}

\section{Conclusions}

\is{
In this work we studied the \velajr\ SNR through the modeling of its broadband non-thermal emission aiming to explain the observed spectral energy distribution together with spatial variations of intensity and spectral parameters. One of the main questions which we try to address is the nature of observed X-ray filaments which feature fast decrease of the intensity and strong spectral softening towards the interior of the remnant. These filaments are usually explained by fast synchrotron cooling in \rob{a} strongly amplified magnetic field of $\sim\,100\ \mu\mathrm{G}$, but can also be caused by the fast decrease of the magnetic field strength downstream of the shock due to the magnetic field damping. The latter effect does not require as strong a magnetic field downstream of the shock and therefore might be a more plausible explanation for the X-ray filaments observed in the \velajr\ SNR. Indeed, our simulations favor the leptonic scenario suggesting that the observed gamma-ray emission is produced predominantly via IC scattering of relativistic electrons accelerated at the shock of the remnant implying that the downstream magnetic field is $8-13\,\mu$G. The contribution of the gamma-ray emission produced in hadronic interactions appears to be negligible.

We consider two different descriptions for the magnetic field profile downstream of the shock, for the damped and transported magnetic fields. In both cases the softening of the X-ray spectrum as well as the X-ray intensity profile is determined by the decrease of the magnetic field strength with the distance from the shock. The synchrotron cooling plays a less important role due to the low value of the magnetic field strength. 
The projection effect in the spherically symmetric case smears the spatial variation of both intensity and spectral parameters resulting in a similar trend but much weaker variation. The observed depth of the decrease of the X-ray intensity cannot be reproduced even for the ideal case when all the emission is coming from the location of the shock, strongly indicating a significant deviation from spherical symmetry. We constructed an asymmetric model for the damped field scenario by assuming that the X-ray emission is enhanced within \rob{a} cone with the opening angle $\theta = 15^\circ$ which roughly corresponds to the the size of the observed X-ray filament. It is interesting that in this scenario both the intensity profile and the spectral variation can be explained quite well. The reason for the enhanced emission inside this cone can be more efficient particle injection due to the higher ambient density. This idea seems to be in agreement with the recently detected correlation of the HI gas distribution with bright TeV emission regions which also correspond to local brightenings at X-ray energies. Moreover, if the enhanced X-ray emission is really caused by the more efficient electron injection, this would also cause the enhancement of the TeV emission in a similar fashion. Remarkably, the choice of the cone parameters tuned for the fit of the X-ray properties also results in a much better fit of the azimuthally averaged TeV intensity profile \isr{as compared to the spherically symmetric model for the damped field}. As a result, for the asymmetric case both X-ray and TeV intensity profiles can be reproduced within the damped field scenario. \isr{The choice of the cone parameters is also perfectly in agreement with the observed azimuthal profiles at X-ray and TeV energies.}

In our simulations we use two different forms of the spatial diffusion coefficient (Bohm diffusion and Kolmogorov-type diffusion) which determines the shape of the cut-off of the particle spectrum. These two parametrizations can be considered as extreme cases defining the range of possible scenarios. Independent simulations including the solution of the transport equation for magnetic turbulence driven by Alfv\'en waves result in a particle spectrum with a cut-off slower than for the Bohm diffusion but faster than for the Kolmogorov-type diffusion. The detected TeV and X-ray spectra favor different types of diffusion. The X-ray spectrum is better reproduced by the Kolmogorov-type diffusion. The Bohm diffusion implies a much faster cut-off which is also in contradiction with the recent detection of hard X-rays from the northern rim of the remnant. However, the shape of the cut-off at TeV energies slightly favors the Bohm diffusion.   

All our models strongly underpredict radio emission from the remnant. This might indicate that we do not account for some other processes which can take place in the SNR and be responsible for the radio emission. One of the possibilities can be stochastic re-acceleration of electrons at the turbulence in the immediate downstream region of the shock. However, it is also possible that a part of the radio emission coincident with \velajr\ might be associated with the Vela SNR which is located in the foreground. In that case the real radio flux from the remnant will be lower.

Further theoretical studies of the \velajr\ SNR would strongly benefit from future observations with CTA which will be able to offer higher sensitivity and better spatial resolution at VHEs. \isr{Spatially resolved spectroscopy at TeV energies might strongly constrain the distribution of the magnetic field downstream the shock and either confirm or discard the magnetic damping scenario. While in the transported field scenario the softening of the X-ray spectrum is due to the change of the maximum electron energy and hence shouuld also be reflected at TeV energies, in the damped field scenario the cut-off energy of the synchrotron spectrum changes due to the decrease of the magnetic field and thus the spectral variation at TeV energies (if any) does not have to correlate with X-rays. Therefore,} spatially resolved comparison of the TeV and X-ray data is crucial for the understanding of the physical processes which take place in this SNR. 

}



\bibliographystyle{aa} 
\bibliography{velajr_damp.bbl}





\end{document}